\documentclass[12pt]{iopart}
\usepackage{bm}
\usepackage{epsfig}
\usepackage{amssymb}
\usepackage{graphicx}
  \expandafter\let\csname equation*\endcsname\relax
  \expandafter\let\csname endequation*\endcsname\relax 
\usepackage{amsmath}
\usepackage{bbold}
\newcommand{\ket}[1]{|#1\rangle}
\newcommand{\bra}[1]{\langle#1|}  
\begin{document}
\title{Entanglement concentration for two atomic ensembles using an effective atom-light beamsplitter}
\author{R Tatham and N Korolkova}
\address{School of Physics and Astronomy, University of St Andrews, Fife, Scotland}
\ead{\mailto{rt264@st-andrews.ac.uk}, \mailto{nvk@st-andrews.ac.uk}}
\begin{abstract}
We present a protocol for increasing the entanglement between two entangled atomic ensembles based on applying an approximate atom-light beamsplitter transformation to both ensembles. The effective asymmetric atom-light beamsplitter is created via a double-pass quantum non-demolition interaction between polarized light and a spin polarized atomic ensemble, derived from the linearised dipole interaction. The entanglement concentration protocol itself uses the procrustean method, similar to that first devised for light by Browne \etal 
[2003 \PR A {\bf 67} 062320] and includes photon counting after the interaction as the required non-Gaussian element. We calculate the output logarithmic negativity in this scheme and show that entanglement between macroscopic ensembles can be increased with probabilities comparable with those for the light scheme even if on-off detectors of low efficiency are used.
\end{abstract}
\submitto{\jpb}
\maketitle

\section{Introduction}

Quantum communication schemes, from quantum key distribution to teleportation and dense coding, offer better ways to exchange information in a network. Experiments into these areas are abundant and have yielded lots of successes over small distances. However, over larger distances, optical losses, phase diffusion, and mixing with thermal states cause the signals to decohere over some finite transmission length. In fact, the error probability scales exponentially with the length of the channel.

The most obvious way to overcome this would be to create a series of signal amplifiers to be spaced out between the source and destination where the signal could be stored onto a quantum memory device and read out again. Such quantum repeater protocols have been devised \cite{Briegel1998} with slight variances between schemes, but are  in general reliant on entanglement distillation procedures and quantum memory devices that can be used to store a signal for a brief period of time.

Due to the long lifetimes of atomic states, the most promising protocol so far uses the coherent spin states of cesium atoms to serve as a memory device. The quantum signal to be transmitted is contained in the polarization state of the incoming light mode and this information is written onto the macroscopic coherent spin state of the atomic ensemble. Julsgaard \etal \cite{Julsgaard2001}
successfully showed that the coherent spin states of two atomic ensembles could become entangled in such a way as to be analogous to the two mode squeezed state for light, written in the number basis 
\begin{equation}
\label{tmss}
\left| \textrm{TMSS}\right\rangle =\sqrt{1-\lambda^2}\sum^\infty_{n=0}\lambda^n\ket{n}_1\ket{n}_2
\end{equation}
where $n$ is the photon number and $\lambda$ quantifies the reduction (squeezing) of the quantum uncertainty of the global state. We will discuss the particular meaning of these quantities in the case of atomic ensembles in Section~\ref{concentration}. The better entangled the ensembles, the better the outcome of the quantum repeater protocol. However the entanglement of two atomic ensembles is at present limited and yields teleportation fidelities of $<0.6$ \cite{Julsgaard2001}. For this reason, the option to increase entanglement in the ensembles by local operations would be greatly desired.

Entanglement distillation \cite{Bennett,Deutsch} is a large field of research, and many different protocols have been devised for continuous variable entanglement concentration \cite{David12,Marek,Hage1}. In recent years there has also been experimental success in distilling light states \cite{Ourjoumtsev2007,HageNature,Dong} but little is known about the possibility of distilling entanglement in atomic ensembles.

We here demonstrate a theoretical protocol for increasing the entanglement between two atomic ensembles. Each atomic ensemble is made to interact with an incoming polarized light mode. Both emerging light modes are then redirected back into their respective ensembles and reemerge to be measured for reoriented photons using a photon counting technique implemented by on-off detectors. If both detectors respond in the affirmative, the entanglement between the two atomic ensembles can be shown to have increased. This scheme is similar in nature to one already devised for light (described in \cite{Browne2003,Eisert2004} and the first demonstration reported in \cite{Ourjoumtsev2007}). Note that the experimental schemes in \cite{Marek,Hage1,HageNature,Dong} present entanglement {\it purification}. That is, they begin with a mixed entangled state containing some non-Gaussian noise and aim at reducing that noise thus increasing the degree of entanglement. In contrast, the original procrustean scheme \cite{Ourjoumtsev2007,Browne2003,Eisert2004}, the schemes based on the Kerr nonlinearity \cite{David12,Radim} and the atomic scheme presented here deal with entanglement {\it concentration}. That is, the aim is the increase of entanglement content in an initially pure Gaussian partially entangled state. 

In Section \ref{system} the quadrature system and interaction used in the scheme are described. In Section \ref{concentration} the entanglement concentration is shown with losses taken into account in Section \ref{losses}. In Section \ref{discuss} we conclude with discussion of the performance of the scheme.

\section{The system and interactions}
\label{system}

The standard QND Hamiltonian couples one of the two effective quadrature field operators of the strongly polarized light mode (re-scaled polarization variables) with one of the effective quadratures of the spin-polarized atomic ensemble, that is with a certain component of the re-scaled collective spin operator (for review see \cite{double-polzik}). The polarization variables are defined as follows. We assume the light beam to be strongly polarized in the $x$-direction so that the actual polarization variable, the Stokes operator $\hat{S}_x$, can be replaced by its expectation value $\left\langle S_x\right\rangle$. The Stokes operators can then be defined, e.g. for a pulse travelling in the z-direction, by
\begin{eqnarray}
 \hat{
S}_x=\frac{c}{2}\int^T_0\left( \hat{a}^\dagger_x\hat{a}_x-\hat{a}^\dagger_y\hat{a}_y\right)d\tau \approx\left\langle S_x\right\rangle=\frac{A_x}{2},
\\ \hat{S}_y=\frac{c}{2}\int^T_0\left(\hat{a}^\dagger_x\hat{a}_y+\hat{a}^\dagger_y\hat{a}_x\right)d\tau,
\\ \hat{S}_z=\frac{c}{2i}\int^T_0\left(\hat{a}^\dagger_x\hat{a}_y-\hat{a}^\dagger_y\hat{a}_x\right)d\tau
\end{eqnarray}
where $\hat{a}_{x,y}\equiv\hat{a}_{x,y}\left(z,t\right)$ are the annihilation operators for photons linearly polarized in the $x$- and $y$- directions respectively with $\left[\hat{a}_i\left(z,t\right),\hat{a}^\dagger_j\left(z',t\right)\right]=\delta_{ij}\delta\left(z-z'\right)\delta\left(t-t'\right)/c$ and $A_x$ is the real expectation value of $\hat{a}_x$ and $\hat{a}^\dagger_x$ when highly polarized. $T$ is the interaction time and $\tau=t-z/c$. With this in mind we can define continuous variable quadratures for the light state $\hat{X}_L$ and $\hat{P}_L$ as
\begin{equation}
 \hat{X}_L=\frac{\hat{S}_y}{\sqrt{\left\langle S_x\right\rangle }},\quad\quad\hat{P}_L=\frac{\hat{S}_z}{\sqrt{\left\langle S_x\right\rangle }}.
 \end{equation}
Here $\left[\hat{X}_L,\hat{P}_L\right] =i$ as with conjugate position and momentum and we have set $\hbar=1$. Note that these quadratures describe the quantum polarization of light and are not the usual amplitude and phase quadratures.

A similar description for atoms is possible, also using appropriate, re-scaled ``quadratures''. Let us first define the relevant variables. The collective angular momentum of the atomic ensembles $\hat {\bf J}$, which has components $\hat J_j, j=x,y,z$ obey the same SU(2) algebra as the Stokes operators, which makes it particularly easy to map quantum states of both systems onto each other. We are interested in increasing the entanglement of two macroscopic atomic ensembles, for example of two macroscopic ensembles of $N_a$ cesium atoms at room temperature with a ground state degeneracy as used in Julsgaard \etal \cite{Julsgaard2001}. There, in a homogenous magnetic field the cesium atoms were pumped into the $\left|F=4,m_F=4\right\rangle$ state in the first cell and $\left|F=4,m_F=-4\right\rangle$ in the second cell to form coherent spin states oriented in the $+x$ and $-x$ directions respectively. In this way, the collective angular momentum in the $x$-direction can also be replaced by the expectation value and $\left\langle \hat{J}_{x1}\right\rangle =-\left\langle \hat{J}_{x2}\right\rangle $ with subscripts $1,2$ representing ensembles $1$ and $2$. In the language of density operators $\hat{\sigma}_{\mu,\nu}=\left|\mu\right\rangle\left\langle \nu\right|$ the operators can be written as
\begin{equation}
\hat{J}_x=\frac{N_a}{2}\sum_{m_F} m_F\hat{\sigma}_{m_F,m_F},
\end{equation}
\begin{equation}
\hat{J}_y=\frac{N_a}{2}\sum_{m_F}C\left(F,m_F\right)\left(\hat{\sigma}_{m_F+1,m_F}+\hat{\sigma}_{m_F,m_F+1}\right),
\end{equation}
\begin{equation}
\hat{J}_z=\frac{N_a}{2i}\sum_{m_F}C\left(F,m_F\right)\left(\hat{\sigma}_{m_F+1,m_F}-\hat{\sigma}_{m_F,m_F+1}\right)
\end{equation}
where
\begin{equation}
C\left(F,m_F\right)=\sqrt{F\left(F+1\right)-m_F\left(m_F+1\right)}.
\end{equation}

As the atomic ensemble is spin-polarized, conjugate position and momentum quadratures for atoms (subscript $A$)  can be defined as (e.g.\cite{double-polzik},\cite{Duan2000}):
\begin{equation}
 \hat{X}_{A}=\frac{\hat{J}_{y}}{\sqrt{\left\langle J_{x}\right\rangle}},\quad\quad\hat{P}_{A}=\frac{\hat{J}_{z}}{\sqrt{\left\langle J_{x}\right\rangle}}. 
\end{equation}

The interaction between light and atoms is represented by the linearised dipole interaction with far-off detuning (off-resonant interaction):
\begin{equation}
 \hat{H}=\sum_j -\bf{d}_j\cdot\bf{E}(\bf{R}_j)
\end{equation}
where $\bf{d}_j=-e\bf{r}_j$ is the dipole operator for the $j$th atom and $\bf{R}_j$ is the location of the $j$th atom. If, for example, the polarized light propagates in the $z$-direction through an atomic ensemble, the linearised interaction Hamiltonian can be written as
\begin{equation}
\label{interac}
 \hat{H}=a\int^T_0\hat{S}_z(t)\hat{J}_z(t)dt\approx\kappa\hat{P}_L\hat{P}_A
\end{equation}
where $a$ is a coupling constant and $\kappa=a\sqrt{\left\langle S_x\right\rangle T\left\langle J_x\right\rangle} $. Any higher order coupling terms are negligible if the laser beam is far detuned from the transition frequencies.

\Eref{interac} is an example of a Quantum Non-Demolition (QND) Hamiltonian which would result in a phase shift in the $\hat{X}_L$ quadratures of the light by $+\kappa\hat{P}_A$ as it passes through the atoms. There is a corresponding back action on the distribution of spins in the atomic ensemble, represented by a phase shift of $+\kappa\hat{P}_L$ in the $\hat{X}_A$ quadrature of the atomic ensemble. Physically, the interaction causes the polarization of the light to rotate about the axis of propagation, dependant on the quadrature distribution of the atoms. The back action effect on the atoms is to rotate the macroscopic spin state around the axis of propagation. 

Whereas a single pass of a light pulse through an atomic medium corresponds to the simple QND interaction described above, multiple passes open the possibility for a larger design freedom for the effective Hamiltonian, as for each pass a particular form of the underlying QND interaction can be adjusted (see e.g. \cite{polzik-multiple}). A double pass scheme \cite{sherson} can be used for the generation of polarization squeezed light by optical Faraday rotation. Another double pass scheme was suggested and thoroughly studied in the context of quantum memory for light modes based on macroscopic atomic ensembles at room temperature, as well as for the generation of entanglement between light and atoms  \cite{mushik}. There, the interaction Hamiltonian has been shown to include two main parts, one equivalent to the beamsplitter interaction, and the other to two-mode squeezing. Depending on the geometry of the setup, either of the two underlying dynamics can be selected. In this paper we exploit the fact that the double-pass Hamiltonian can approximate, with high fidelity, an actual beamspitter transformation, the quality of the approximation being dependent not only on the interaction strength, but also on the particular quantum states of the interacting modes \cite{approxBS-tatham}. Tuning the interaction parameter, a highly asymmetric atom-light beamsplitter can be realized. In what follows, we describe the entanglement concentration scheme based on such an atom-light beamsplitter and analyse the performance of the scheme.

\section{Entanglement Concentration}
\label{concentration}

In the continuous variable regime, to increase the entanglement between two quantum objects with Gaussian quadrature distributions a non-Gaussian element is required. The measurement process offers this opportunity. Analogous to the photon subtraction schemes examined in \cite{Kitagawa2006} and \cite{Ourjoumtsev2007} for increasing the entanglement in two mode squeezed states of light, a photon count heralds an increase in entanglement between two macroscopic atomic ensembles. 

\begin{figure}
\begin{center}
\includegraphics[width=13cm]{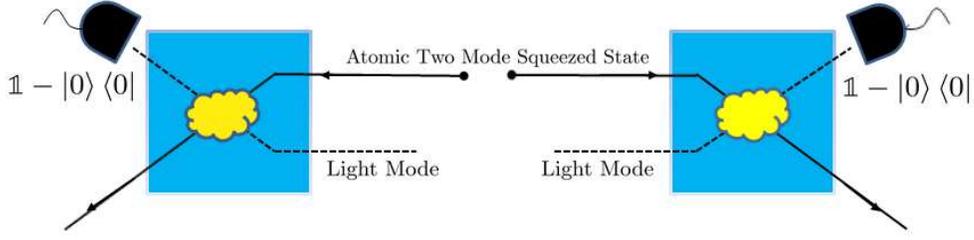}
\caption{Procrustean entanglement concentration for atomic ensembles: the principle. The entanglement concentration protocol works similar to the light scheme \cite{Browne2003,Eisert2004}. The two ensembles are entangled like a TMSS. Then $x$-polarized light interacts with the atoms (number basis $\left|0\right\rangle$) and non-Gaussian measurements are done to see whether the polarization of the photons has altered. A positive response from both detectors heralds an increase in entanglement between the atomic ensembles. The two square boxes symbolize the effective atom-light beamsplitter based on the double QND interaction.}
\end{center}\label{Fig1}
\end{figure}

\begin{figure}
\begin{center}
\framebox{
\includegraphics[width=10cm]{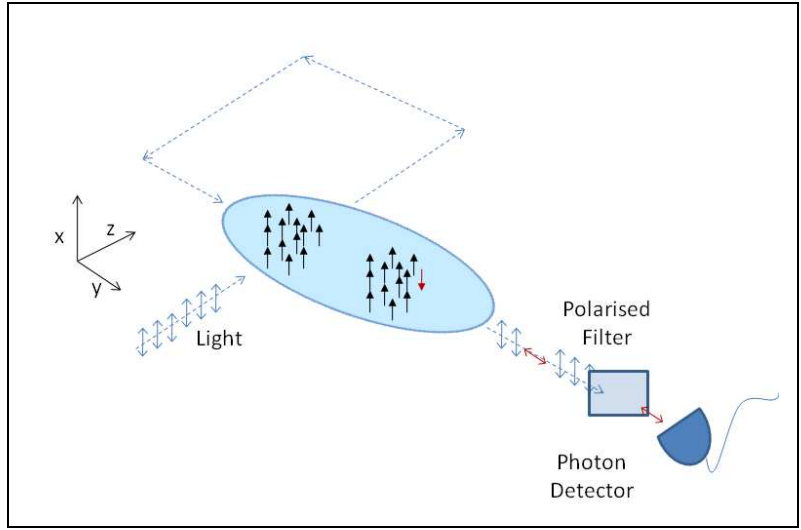}
}
\caption{Procrustean entanglement concentration for atomic ensembles: details of atom-light interactions. Light is strongly polarized in the $x$-direction. It interacts with the spin-polarized atomic ensemble in a double-pass interaction equivalent to the effective atom-light beamsplitter which probabilistically accomplishes photon subtraction in a light beam. A $y$-polarized photon detection at the output implies that the atomic entanglement has increased.}
\end{center}\label{Fig2}
\end{figure}

In the case of the two mode squeezed state for light, the scheme is fairly simple, the two correlated beams are sent to separate highly transmissive beamsplitters, upon which they combine with another mode (usually the vacuum). There is a small probability that a photon will be subtracted from the main beam and proceed to be detected by a photon counter. If the counters at both beamsplitters register the presence of photons, then the entanglement of the two mode squeezed state increases. We wish for a similar ``photon subtraction'' scheme for our light and atoms in order to increase the entanglement between the two atomic ensembles as represented in Figure 1. 

The two atomic ensembles are prepared in an entangled two mode squeezed state as described in \cite{Julsgaard2001} . 
Previous works \cite{Julsgaard2001,Duan2000,polzik} have detailed the methods for entangling the collective atomic spins of the atomic ensembles using equations of the form \eref{interac}. By sending a polarized light mode through the atomic ensembles and taking a homodyne measurement, it is possible to collapse the atomic states into an entangled two mode squeezed state with variance $\Delta\left(\hat{X}_{A1}-\hat{X}_{A2}\right)=\Delta\left(\hat{P}_{A1}+\hat{P}_{A2}\right)=e^{-2r}$ where $r$ is dependent on $\kappa$ and given by
\begin{equation}
\label{squee}
 r=\frac{1}{2}\ln\left(1+2\kappa^2\right).
\end{equation}
Further, it is assumed that any displacement of the joint atomic quadrature distributions from the centre in phase space caused by the entanglement process itself are small so that the constraints on the beamsplitter-like interaction still hold \cite{approxBS-tatham}. For ease of calculation, the atomic state is represented in the number basis as in \Eref{tmss} where $\lambda=\tanh(r)$ is the squeezing parameter, dependent on the interaction strength $\kappa$ between a light mode and atomic ensemble during the initial entanglement process via the relation \eref{squee}. The number basis is considered to be the basis generated by the annihilation operator $\hat{a}_A=\hat{X}_A+i\hat{P}_A$ up to normalisation. On consideration the number $n$ represents the number of atoms in the upper excited spin state in the basis of the $\hat{J}_x$ operator. That is, $n=1$ corresponds to the superposition of all possible combinations of atomic spins of the atoms in the ensemble for which a single atom is excited.

If the light is initially prepared in a symmetric Gaussian state centred at $X_L=0$, $P_L=0$ then the light can be considered to be in a vacuum state. That is, there are no photons polarized in the $\hat{S}_y$ or $\hat{S}_z$ directions despite a steady base stream of linearly $x$-polarized photons so that $\left\langle S_y\right\rangle =\left\langle S_z\right\rangle =0$. In the number  basis generated by operators $\hat{S}_+=\hat{S}_y+i\hat{S}_z=\sqrt{\left\langle S_x\right\rangle }(\hat{X}_L+i\hat{P}_L)$ and $\hat{S}_-=\hat{S}^\dagger_+$ it is clear that a Fock state is simply the number of photons polarized linearly in the $y$-direction.

In the number basis, the functionality of the photon subtraction scheme can be readily explained. If, after the beamsplitter-like interaction, one or more photons are detected behind the polarized filter, then the interaction has rotated the polarization of some photons in the light mode. The corresponding back action on the atomic ensembles is to flip the spin of one or more of the atoms in the system (see Fig.~2). 

The initial state of the light and atoms is given by 
\begin{equation}
 \left|\Psi_I\right\rangle_{LA}=\sqrt{1-\lambda^2}\sum^\infty_{n=0}\lambda^n\ket{n}_1\ket{n}_2\ket{0}_3\ket{0}_4
\end{equation}
where the first and second atomic modes are subscripted $1$ and $2$ respectively and the light vacuum modes are subscripted $3$ and $4$. The light vacuum modes $3$ and $4$ (i.e. modes that are only polarized in the $x$-direction), are sent through the atomic ensembles $1$ and $2$ respectively and made to interact twice with the ensembles via the beamsplitter interaction described in \cite{approxBS-tatham}. That is, in one atom-light ``beamsplitter'' two QND interactions are performed. The first is of the form $\hat{H}_1=\phi\hat{P}_L\hat{X}_A$ and the outgoing light modes are redirected back into the atomic ensembles to interact via a second interaction, $\hat{H}_2=-\phi\hat{X}_L\hat{P}_A$. We use the reasonable approximations on the quadratures that $x_l\approx\sqrt{1-\phi^2}x_l$ and $\left(1-\phi^2\right)x_a\approx\sqrt{1-\phi^2}x_a$ for small $\phi$ with similar conditions on $p_l$ and $p_a$ \cite{approxBS-tatham}. This double-pass scheme then performs the role of a beamsplitter transformation and is treated accordingly in what follows. The state after the beamsplitter interaction can be described by
\begin{eqnarray}
\left| \Psi_F\right\rangle_{LA} =\sum^\infty_{n=0}\sqrt{1-\lambda^2} \lambda^n &\times\sum^n_{k_1,k_2=0} \sqrt{\binom{n}{k_1}\binom{n}{k_2}}\nonumber\\ \times \phi^{k_1+k_2} \left(1-\phi^2\right)^{2n-k_1-k_2}&\times \ket{n-k_1}_1 \ket{n-k_2}_2 \ket{k_1}_3 \ket{k_2}_4
\end{eqnarray}
where $\phi$ is the strength of the interaction.

The outgoing light modes are then passed through a polarized filter to remove the base stream of $x$-polarized photons and allowing only photons whose polarizations have been rotated by the interaction until $y$-polarized to proceed (Fig.~2).
A detector then registers whether $y$-polarized photons are present. As detectors are still relatively inefficient at counting photons, it is instead assumed that the detectors used can, to a high degree of efficiency, detect simply the presence of one or more photons (``on-off detector''). The state thus becomes:
\begin{eqnarray}
\left|\Psi_A\right\rangle_{\textrm{out}}&=\sqrt{\frac{1-\lambda^2}{S}}\sum^\infty_{n=0}\sum^\infty_{u,v=1}\lambda^n\sqrt{\binom{n}{u}\binom{n}{v}}\nonumber\\&\quad\times\phi^{u+v}\left(1-\phi^2\right)^{2n-u-v}\ket{n-u}_1\ket{n-v}_2 \\
\end{eqnarray}
where S is the probability of getting an affirmative measurement at both detectors:
\footnotesize
\begin{eqnarray}
S&=\frac{1-\lambda^2}{1-\lambda^2\left(\phi^2+\left(1-\phi^2\right)^2\right)^2}-\frac{2\left(1-\lambda^2\right)}{1-\lambda^2\left(1-\phi^2\right)^2\left(\phi^2+\left(1-\phi^2\right)^2\right)}\nonumber\\&+\frac{1-\lambda^2}{1-\lambda^2\left(1-\phi^2\right)^4}.	\\
\end{eqnarray}
\normalsize
The amount of entanglement in the two ensembles is quantified by the negativity and logarithmic negativity \cite{Vidal2002} of the state defined by
\begin{equation}
\mathcal{N}(\hat{\rho})=\frac{1}{2}{\textrm{Tr}}\left(\sqrt{(\hat{\rho}^{PT})^2}-\hat{\rho}^{PT}\right)=\frac{\left\|\hat{\rho}^{PT}\right\|-1}{2},
\end{equation}
\begin{equation}
\mathcal{E}_\mathcal{N}(\hat{\rho})=\ln(1+2\mathcal{N}(\hat{\rho}))=\ln(\left\|\hat{\rho}^{PT}\right\|)
\end{equation}
where $\left\|\cdot\right\|$ denotes the trace-norm, i.e. $\sqrt{\hat{\rho}\hat{\rho}^\dagger}$, and $\hat{\rho}^{PT}$ is the partial transpose of the density matrix $\hat{\rho}$. 

It is simple to show that for a two mode squeezed state, the negativity and logarithmic negativity are given as
\begin{equation}
\label{ntmss}
\mathcal{N}(\textrm{TMSS})=\frac{\lambda}{1-\lambda},
\end{equation}
\begin{equation}
\label{etmss}
\mathcal{E}_\mathcal{N}(\textrm{TMSS})=\ln(1+\lambda)-\ln(1-\lambda).
\end{equation}
For projections onto an exact photon state (e.g., for a single photon detection), the negativity and logarithmic negativity can be calculated analytically. However, for on-off type measurements the entanglement measures must be calculated numerically due to the infinite sums over $u$ and $v$. Fortunately, these sums appear to converge quickly and so we can set a reliable truncation point. The calculation is done in a similar way to the calculations of Kitagawa \etal \cite{Kitagawa2006}. Firstly, the density matrix of the state is expanded as
\begin{equation}
\label{calcbegin}
\ket{\Psi_A}_{\textrm{out}}\bra{\Psi_A}=\sum^\infty_{a,b,c,d}\rho_{a,b,c,d}\ket{a}_{1}\bra{c}\otimes\ket{b}_{ 2}\bra{d}
\end{equation}
where
\begin{eqnarray}
\rho_{a,b,c,d}&=\left(_1\bra{a}_2\bra{b}\right)\ket{\Psi_A}_{\textrm{out}}\bra{\Psi_A}\left(\ket{c}_1\ket{d}_2\right) \nonumber\\
&=\frac{(1-\lambda^2)}{S}\sum^\infty_{u,v=1}\lambda^{a+u}\lambda^{c+u}\nonumber\\&\quad\times\sqrt{\binom{a+u}{u}\binom{a+u}{v}\binom{c+u}{u}\binom{c+u}{v}}\nonumber\\&\quad\times\left(\phi\right)^{2(u+v)}\left(1-\phi^2\right)^{2(a+c+u-v)}\delta_{a-b,v-u}\delta_{c-d,v-u}.\\
\end{eqnarray}
The partial transpose of this state is given by
\begin{equation}
\label{statetranspose}
\left(\ket{\Psi_A}_{ \textrm{out}}\bra{\Psi_A}\right)^{PT}=\sum^\infty_{a,b,c,d}\rho_{a,d,c,b}\ket{a}_1\bra{c}\otimes\ket{b}_2\bra{d}
\end{equation}
and the elements are zero unless the total Fock number of the entangled state, $N=a+b=c+d$, is non-zero. This follows from the delta functions in $\rho_{a,d,c,b}$. Operator \eref{statetranspose} is block diagonal and we can write it as a direct sum of each $N$-dependent submatrix:
\begin{equation}
\label{calcend}
\left(\ket{\Psi_A}_{ \textrm{out}}\bra{\Psi_A}\right)^{PT}=\oplus^{\infty}_{N=0}\left(\ket{\Psi_A}_{ \textrm{out}}\bra{\Psi_A}\right)^{PT}(N)
\end{equation}
where $\left(\ket{\Psi_A}_{ \textrm{out}}\bra{\Psi_A}\right)^{PT}(N)$ is the $(N+1)\times(N+1)$ $N^{th}$ submatrix.

\begin{figure}[!bth]
\begin{center}
\includegraphics[width=10cm]{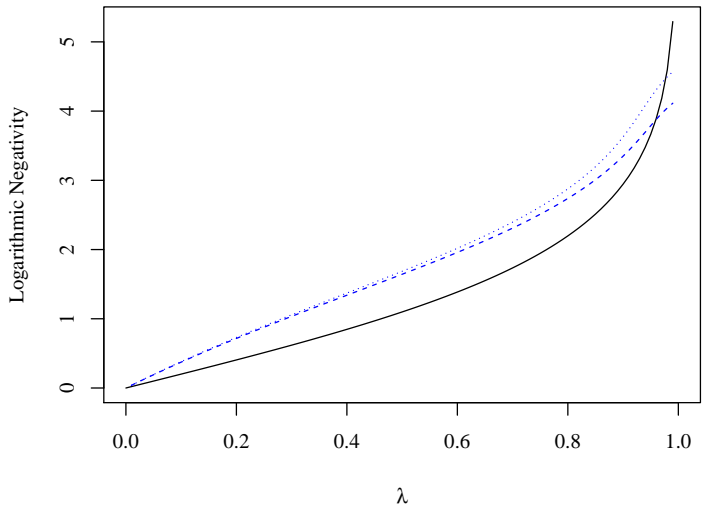}
\end{center}
\caption{\small A plot of the logarithmic negativity against $\lambda$ for (i) standard two mode squeezed state (solid) and for the output of the photon subtracted scheme for the beamsplitter like interaction between light and atomic ensemble when (ii) $\phi=0.1$ (dashed) and (iii) $\phi=0.01$ (dotted). See text for discussion.}
\label{Fig3}
\end{figure}

\begin{figure}
\begin{center}
\includegraphics[width=10cm]{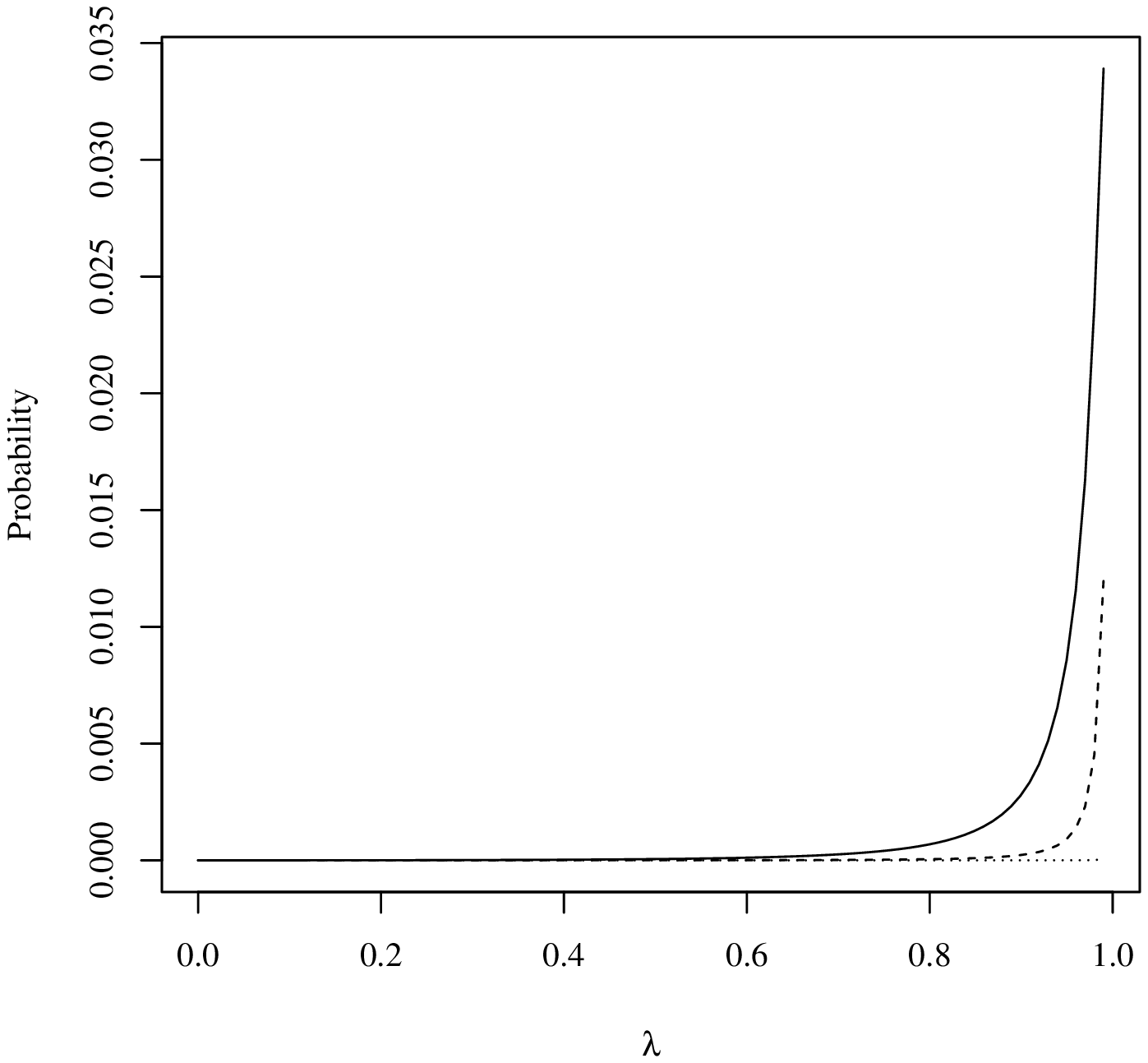}
\end{center}
\caption{\small Probability of success against $\lambda$ for (i) interaction strength $\phi=0.1$ (solid) (ii) $\phi=0.05$ (dashed) and (iii) $\phi=0.01$ (dotted). The probabilities of success are small but comparable with the light scheme.}
\label{Fig4}
\end{figure}

The negativity of the partially transposed state is then computed by numerically diagonalising each block individually to obtain the eigenvalues of each submatrix and adding up the absolute value of all negative eigenvalues. A cut-off, $N_{max}$, must be introduced that is large enough compared to the mean number of excited spins. There is of course a trade-off between $N_{max}$ and the length of time needed to perform the calculation. For the purposes of this calculation, a value of $N_{max}=100$ gave very precise results. That is, at $N_{max}=100$ the numerical values of logarithmic negativity converge to at least 7sf and to increase $N_{max}$ beyond this is not beneficial. The results are shown in Figure~3. 

The entanglement between the two atomic ensembles is increased for all values of initial squeezing except for very high $\lambda$. This increase in $\mathcal{E}_\mathcal{N}$ is not very large but comparable to when light modes are used in place of atomic ensembles \cite{Kitagawa2006}, as are the probabilities shown in Figure~4.
As can be seen, a trade-off is required between interaction strength and probability of success. As interaction strength increases, so does the probability of success, but the validity of the beamsplitter approximation decreases. Note, however, that the beamsplitter approximation has a very high fidelity of approximately 0.99 even for interaction strengths as high as $\phi\approx 0.35$. For moderate initial squeezing, $\mathcal{E}_\mathcal{N}$ is largely unaffected by the interaction strength. As $\lambda$ approaches $\approx 0.95$, the concentration procedure ceases to bring further benefit. However, this only occurs at exceptionally high squeezing of the atoms, which is not experimentally viable.

\section{Modelling Detector Inefficiencies}
\label{losses}
The largest contribution to loss in the photon subtraction scheme for light modes comes from detector inefficiency. For the atomic ensemble scheme the efficiency of the detectors will also play a crucial role. The detectors here have a reduced number of photons to detect due to the polarization filter used to stop the base stream of $x$-polarized photons. The inefficiency of the detector can be modelled as an ideal detector behind a beamsplitter of transmittivity $\eta=\nu^2$. The light mode is combined with a vacuum on a beamsplitter and the vacuum is traced out before the projection measurement is performed.
For a Fock state $\left|k\right\rangle$ combining with a vacuum, this amounts to the transformation
\begin{equation}
\ket{k,0}\rightarrow\sum^{k}_{s=0}\sqrt{\binom{k}{s}}\nu^s\left(\sqrt{1-\nu^2}\right)^{k-s}\ket{s,k-s}.
\end{equation}
Directly before the detection is performed, the state of the density matrix is given by
\begin{eqnarray}
\rho=&\left(1-\lambda^2\right)\sum^\infty_{m,n=0}\lambda^{m+n}\sum^n_{k_1,k_2=0}\sum^m_{j_1,j_2=0}\sqrt{\binom{n}{k_1}\binom{m}{j_1}\binom{n}{k_2}\binom{m}{j_2}}\nonumber\\
&\times\phi^{j_1+k_1+j_2+k_2}\left(1-\phi^2\right)^{2n+2m-j_1-k_1-j_2-k_2}\sum^{k_1}_{s=0}\sum^{j_1}_{t=0}\sum^{k_2}_{y=0}\sum^{j_2}_{z=0}N^{k_1,k_2,j_1,j_2}_{s,y,t,z}\nonumber\\&\ket{n-k_1}_{ 1}\bra{m-j_1}\otimes\ket{n-k_2}_{ 2}\bra{m-j_2}\otimes\ket{s}_{ 3}\bra{t}\otimes\ket{y}_{ 4}\bra{z}
\end{eqnarray}
where
\begin{eqnarray}
N^{k_1,k_2,j_1,j_2}_{s,y,t,z}=&\sqrt{\binom{k_1}{s}\binom{j_1}{t}\binom{k_2}{y}\binom{j_2}{z}}\left(\sqrt{1-\nu^2}\right)^{j_1+k_1+j_2+k_2-s-t-y-z}\nonumber\\&\times\nu^{s+t+y+z}\delta_{k_1-s,j_1-t}\delta_{k_2-y,j_2-z}.
\end{eqnarray}
Light modes $3$ and $4$ are subsequently measured for the presence or absence of photons using the operator $\left(\mathbb{1}-\ket{0}\bra{0}\right)$ and traced out. The density matrix of the two remaining atomic modes can then be described by

\begin{eqnarray}
\rho_{\textrm{out},\eta}=&\sum^\infty_{m,n=0}\left(1-\lambda^2\right)\lambda^{m+n}\sum^{\min(m,n)}_{k_1,k_2=0}\sqrt{\binom{n}{k_1}\binom{m}{k_1}\binom{n}{k_2}\binom{m}{k_2}}\phi^{2\left(k_1+k_2\right)}\nonumber\\&\times\left(1-\phi^2\right)^{2n+2m-2k_1-2j_1}\left[1-\left(1-\eta\right)^{k_1}-\left(1-\eta\right)^{k_2}+\left(1-\eta\right)^{k_1+k_2}\right]\nonumber\\&\ket{n-k_1}_{ 1}\bra{m-k_1}\otimes\ket{n-k_2}_{ 2}\bra{m-k_2}
\end{eqnarray}

and the probability of success, taking into account detector losses, is given by
\begin{eqnarray}
S_{\eta}&=\frac{1-\lambda^2}{1-\lambda^2\left[\phi^2+\left(1-\phi^2\right)^2\right]^2}\nonumber\\&\quad-\frac{2\left(1-\lambda^2\right)}{1-\lambda^2\left[\phi^2\left(1-\eta\right)+\left(1-\phi^2\right)^2\right]\left[\phi^2+\left(1-\phi^2\right)^2\right]}\nonumber\\&\quad+\frac{1-\lambda^2}{1-\lambda^2\left[\phi^2\left(1-\eta\right)+\left(1-\phi^2\right)^2\right]^2}.\\
\end{eqnarray}
The effect that detector inefficiency has on the logarithmic negativity of the atomic ensembles can be shown in Figure 5.

\begin{figure}[!htb]
\begin{center}
\includegraphics[width=10cm]{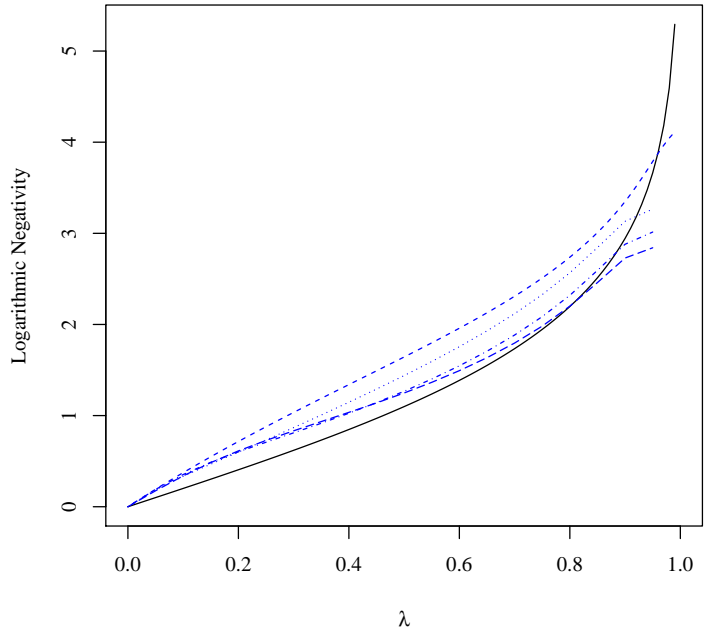}
\caption{Dependence of the logarithmic negativity on the efficiency of the detectors. The solid line depicts the initial two mode squeezed state of the entangled atomic ensembles. Then the logarithmic negativity is shown for $\eta=1$ (dashed line), $\eta=0.8$ (dotted line), $\eta=0.5$ (dash-dot), and $\eta=0.2$ (long dash). The interaction strength is $\phi=0.1$.}
\end{center}
\label{Fig5}
\end{figure}

As expected, the entanglement concentration becomes less pronounced for low detector efficiency. The positive message is that even for efficiencies as low as $\eta=0.2$ there is still a range of $\lambda$ values for which entanglement is increased (although the probability of success is quite low in this case). This range is experimentally accessible and so it is good news that the entanglement concentration protocol is more robust against imperfections.

\section{Conclusions}
\label{discuss}
In this paper, we have employed a beamsplitter-like QND interaction between light and atomic ensembles based on the linearised dipole interaction between strongly polarized light and atomic levels for increasing the entanglement between two atomic ensembles. The entangled atomic ensembles in the initial two mode squeezed state interact with a highly polarized light mode in a quantum vacuum state via effective atom-light beamsplitter. The output light modes are subsequently detected using on-off detectors.  This is analogous to the procrustean entanglement concentration scheme \cite{Browne2003,Eisert2004} based on photon subtraction for distilling entanglement in light modes. Similar to their scheme, in our protocol detector clicks at both light outputs herald the successful ``photon subtraction'' (spin-flip in the atomic ensembles) and thus successful entanglement distillation. To assess the performance of the atomic entanglement distillation scheme, we have calculated the logarithmic negativity for the output quantum state of the two atomic ensembles and shown that it can increase indicating that the entanglement concentration procedure has been successful. The probability of success is very small but comparable to the probabilities for the corresponding light schemes that have been demonstrated experimentally \cite{Ourjoumtsev2007}. The resulting atomic states are non-Gaussian and it remains to be seen whether they can be used for any teleportation procedures without devising a way to re-Gaussify the system. Use of the beamsplitter approximation allows us to closely mimic the entanglement concentration scheme that exists for light and to get an idea of how the light-atom protocol performs in comparison. The next step would be to remove the approximation and to assess the scheme for stronger interactions but a different approach has to be applied in this case.

\ack
The research has been supported by the EU STREP project COMPAS FP7-ICT-2007-C-212008 under the FET-Open Programme, by the Scottish Universities Physics Alliance (SUPA) and by the Engineering and Physical
Sciences Research Council (EPSRC).

\section*{References}

\end{document}